# Autofib Redshift Survey: I – Evolution of the Galaxy Luminosity Function

Richard S. Ellis[1], Matthew Colless[2], Tom Broadhurst[3], Jeremy Heyl[1,4] and Karl Glazebrook[1]
[1] *Institute of Astronomy, Madingley Road, Cambridge CB3 0HA*
[2] *Mount Stromlo and Siding Spring Observatories, The Australian National University, Weston Creek, ACT 2611, Australia*
[3] *Department of Physics & Astronomy, Johns Hopkins University, Baltimore MD 21218, USA*
[4] *Lick Observatory, Board of Studies in Astronomy and Astrophysics, University of California, Santa Cruz, CA 95064, USA*

**ABSTRACT**
We present a detailed determination of the restframe $B$-band galaxy luminosity function (LF) as a function of redshift and star formation activity from $z=0$ to $z \simeq 0.75$. The dataset used for this purpose is a combined sample of over 1700 redshifts spanning a wide range in apparent magnitude, $11.5 < b_J < 24.0$, which we term the Autofib Redshift Survey. The sample includes various earlier magnitude-limited surveys constructed by our team as well as a new survey of 1026 redshifts measured for galaxies at intermediate magnitudes. Spectral classifications, essential for estimating the $k$-corrections and galaxy luminosities, are accomplished via cross-correlation with Kennicutt's library of integrated galaxy spectra. The various overlapping surveys in the sample enable us to assess the effects of redshift incompleteness. We demonstrate that uncertainties in classification and those arising from incompleteness do not seriously affect our conclusions. The large range in apparent magnitude sampled allows us to investigate both the nature of the LF at low redshift ($z < 0.1$) and possible evolution in its shape to $z=0.75$. We find that earlier bright surveys have underestimated the absolute normalisation of the LF. Because the shape of the local LF does not change with the survey apparent magnitude limit, it seems unlikely that the local deficiency arises from an underestimated population of low luminosity galaxies. Furthermore, surface brightness losses cannot be significant unless they conspire to retain the LF shape over a variety of detection thresholds.
Our data directly demonstrates that the $B$-band LF evolves with redshift. This evolution is best represented as a steepening of the faint-end slope of the LF, from $\alpha \simeq -1.1$ at low redshift to $\alpha \simeq -1.5$ at $z \simeq 0.5$. Using [O II] emission as an indicator of star formation activity, we show that the LF of quiescent galaxies has remained largely unchanged since $z \simeq 0.5$, whereas the luminosity density of star-forming galaxies has declined by nearly a factor of 2. The steepening of the overall LF with lookback time is of the form originally postulated by Broadhurst et al. (1988) and is a direct consequence of the increasing space density of blue star-forming galaxies at moderate redshifts.

## 1 INTRODUCTION

The detailed characterisation of the luminosity function (LF) of field galaxies is an important extragalactic question. Notwithstanding several controlled redshift surveys of field galaxies in recent years (Kirshner et al. 1978, Peterson et al. 1985, Loveday et al. 1992), some uncertainty clearly remains in both the absolute normalisation of the LF, $\phi^*$ (cf. Maddox et al. 1990), and the faint end slope, $\alpha$ (Davies 1990, McGaugh 1994). A further important issue is the nature of any dependences of these quantities on morphological type. A steep faint end slope of the LF is a natural consequence of hierarchical models of galaxy formation seeded at early times by cold dark halos (Lacey et al. 1992, Kauffmann et al. 1994, Cole et al. 1994). Improved observational constraints on these models are required.

Our present knowledge of the field galaxy LF comes primarily from redshift surveys limited at $B \approx 17$. Although some of these samples (like the Stromlo-APM and CfA surveys) are extensive, they are not optimally designed to address issues concerning the faint end slope. Their main value has been in defining very precisely the value of $M^*$, verifying that the Schechter (1976) formula is an appropriate representation and providing limited constraints on the form of the LF for $M_B < -13 + 5\log h$ (where $h$ is Hubble's constant in units of 100 km s$^{-1}$ Mpc$^{-1}$). At $B\approx17$, a dwarf galaxy with $M_B = -14 + 5\log h$ can barely be detected beyond the Virgo cluster. Even in panoramic surveys, the volumes probed to this apparent magnitude limit are insufficient to constrain the abundance of such dwarf galaxies. Small local volumes may also be unrepresentative. A further problem with intermediate depth surveys is that the photometric data on which many are based are either not well-defined or are insufficiently deep in their surface brightness limit to reveal possible low surface brightness systems which may dominate the faint LF (McGaugh 1994, Ferguson & McGaugh 1995).

The contribution of dwarf galaxies may be crucial to understanding analyses of deeper ($B>21$) surveys of cosmological importance and, in particular, in quantifying the nature of any faint excess in the galaxy counts (Ellis 1993). Even a minor change in $\alpha$ can produce a dramatic increase in the expected number of $B>21$ galaxies since the faint end of the LF contributes to the number counts with a steep Euclidean slope (Kron 1980, Phillipps & Driver 1995). A related issue here is the normalisation of the local LF. Galaxy counts at intermediate magnitudes $17<B<21$ ( Heydon-Dumbleton et al. 1989, Maddox et al. 1990) present a puzzlingly steep slope. If these data are correct and evolution at such bright magnitudes is discounted, possibly $\phi^*$ may not be well-determined. An upward revision by a factor 2 would reduce the faint excess brighter than $B\approx21$-22 and explain photometric colour and redshift distributions which both match no evolution expectations (Metcalfe et al. 1995a).

Although one motivation for deeper spectroscopic surveys is the need to clarify these uncertainties in the local LF, the main goal for the fainter surveys done to date has been to search for evolution in the LF (see Koo & Kron 1992 and Ellis 1993 for a review of these efforts). Spectroscopic surveys consisting of 100–300 galaxies in strict magnitude-limited samples fainter than $B=21$ have been published by Broadhurst et al. (1988, hereafter BES), Colless et al. (1990, 1993), Lilly et al. (1991,1995) and Cowie et al. (1991). A consistent picture has emerged from these surveys. Notwithstanding the apparent excess of faint galaxies, the redshift distributions reveal no unexpected high or low redshift tails. To first order the $N(z)$ distributions results are compatible with evolution in galaxy number density, rather than in the luminosity scale. Broadhurst et al. claim a rising fraction of star forming galaxies displaying intense [OII] emission but the validity of this result, the only direct evidence for evolution in the population, relies on understanding the various aperture and $k$-correction biases (cf. Koo et al. 1993).

For reasons of observing efficiency, the deep spectral surveys consist of samples restricted to lie within narrow apparent magnitude ranges. This precludes any *direct* estimation of the LF as a function of redshift. For example, although Broadhurst et al. (1988) were able to demonstrate the redshift distribution of their faint survey was consistent with a LF whose faint end slope steepens with increasing redshift (see their Fig. 8), they were not able to observe such steepening directly in their data. The effect proposed by Broadhurst et al. would produce an effective increase in the number density of luminous galaxies at around $M^*$ (and hence the excess counts) *without* distorting the redshift distribution from its no-evolution expectation. Eales (1993) attempted to combine the various surveys to derive a direct estimate of the LF as a function of redshift, however the inhomogeneity and limited size of the datasets then available precluded very reliable conclusions.

In this series of papers we present the results of a comprehensive new survey, the Autofib Redshift Survey, conducted with the AAT's Autofib fibre positioner (Parry & Sharples 1988). The primary role of the new data is to fill a 'gap' in the coverage of apparent magnitudes in the range $B$=17–21 and to significantly increase the size of the sample out to $B$=22.

The scientific motivation of the survey is two-fold. By extending the local surveys to fainter limits, more rigorous constraints can be provided on the faint end slope and normalisation of the local LF. Secondly, with strategically-constructed samples spanning a wide apparent magnitude range, for the first time we can monitor directly any evolution in the form of the LF with redshift. With a large enough sample it is also possible to check for evolution as a function of spectral class.

Galaxy selection in the $B$ photometric band is advantageous for this large survey not only because it makes optimal use of existing data, but also because it maximises the sensitivity to recent changes in the global star formation rate of galaxies of various kinds. Our survey is directly able to address the long-standing question of the origin of the excess number of $B$-band galaxies. It complements recent work in the $I$-band (Lilly 1993, Lilly et al. 1995) and $K$-band (Cowie 1993, Glazebrook et al. 1995b) whose role is equally important in clarifying longer-term changes in galaxy properties over slightly larger redshift baselines.

This first paper in the series presents the main scientific conclusions of the survey. In Paper II (Heyl et al. 1995) we discuss in more detail the luminosity function of various spectral classes as a function of redshift. Paper III (Broadhurst et al. 1995) discusses the observing strategy and presents the redshift survey catalogue and related quantities for over 1700 galaxies.

The plan of this paper is as follows. In Section 2 we briefly summarise our overall strategy, the incorporation of data from previous surveys, and the new observations conducted with Autofib. In Section 3 we discuss the analysis of the data, including a technique developed to derive $k$-corrections for individual galaxies based on a classification of their spectra, and a simple estimator for deriving the luminosity function in different redshift bins. Section 4 presents the results, including new constraints on the local LF and evidence for evolution in the form of the LF with redshift for the entire sample and for various spectral sub-classes. Section 5 discusses the conclusions of the survey in the context of various explanations proposed for the demise of the faint blue galaxy population.

## 2 THE AUTOFIB REDSHIFT SURVEY

### 2.1 Strategy

The principle goal of the new Autofib survey is to extend the range of galaxy luminosities sampled at moderate redshift by sampling the apparent magnitude–redshift plane inbetween the early $B<17$ surveys and the more recent $20<B<24$ surveys. With this broad coverage of apparent magnitude, a *direct* estimate of the luminosity function (LF) at various redshifts can be obtained. A detailed account of our observing strategy and sample selection will be given in Paper III. Here we briefly summarise the salient points.

The new data consists of 1028 redshifts in 32 pencil beams within two apparent magnitude ranges: $17<b_J<20$ (AF-bright) and $19.5<b_J<22$ (AF-faint). By sampling many different directions rather than a single contiguous area the confusing effects that galaxy clustering may have on the derived LFs can be minimised. The different sampling rates for the various magnitude ranges enable us to make effective use of a limited amount of observing time and populate the apparent magnitude–redshift plane in a well-controlled way.

Table 1 summarises the overall survey characteristics. As well as the new data, we have included the brighter DARS survey (Peterson et al. 1985) and the fainter surveys of BES (Broadhurst et al. 1988), LDSS-1 (Colless et al. 1990, 1993) and LDSS-2 (Glazebrook et al. 1995a). In total our catalogue contains 1701 galaxy redshifts and 3 QSOs. The galaxies have redshifts up to $z=1.108$; the QSOs have $z=1.262$, 1.493 and 1.599. The combined survey consists of 53 pencil beams and spans the apparent magnitude range $b_J=11.5$–24.0. The large number of pencil beams span many widely-separated fields over the entire southern sky thus a very large volume is effectively random-sampled. Paper III in this series (Broadhurst et al. 1995) presents the combined survey catalogue and a field-by-field summary of the selection criteria, sampling rate and redshift completeness.

Details of the photometric selection, observing techniques and spectroscopic analyses for the published data can be found in the relevant references or in Paper III. All galaxy photometry has been reduced to the colour-corrected photographic $b_J \equiv$ Kodak IIIa-J plus GG395 at a limiting surface brightness of $\mu_J=26.5$ mag arcsec$^{-2}$ (Jones et al. 1991). For the new data in the intermediate range observed with Autofib, objects were selected from COSMOS measuring machine scans of sky-limited UK Schmidt plates using a typical threshold of $\mu_J=25.0$ mag arcsec$^{-2}$. This photometry was calibrated with reference to $19<b_J<21$ galaxies in the APM galaxy survey (Maddox et al. 1990) in all cases where the fields overlap, and with the Edinburgh-Durham southern galaxy survey (Heydon-Dumbleton et al. 1989) for the remainder. In producing a uniform photometric catalogue, corrections were made for the different isophotes used in each catalogue (Peterson et al 1985). These corrections are always smaller than $0^m.28$ and thus comparable to the random photometric errors which vary from 0.05-0.15 mag across the catalogue.

Star/galaxy separation for the DARS and BES data was performed by eye. For the fainter LDSS-1 and LDSS-2 surveys *all* objects were observed spectroscopically, and galaxy samples were defined from the spectra obtained. Whereas the penalty of including stars in the deep surveys is small, the additional overhead of this mode of observing at $b_J=17$–20 would be prohibitive. Previous all-inclusive surveys (Tritton & Morton 1984, Colless et al. 1990, 1991, 1993, Glazebrook et al. 1995a) have failed to find a significant extragalactic population of compact sources. In the new data reported here, we therefore relied on the COSMOS star-galaxy classification algorithm, making additional visual checks of each selected target prior to undertaking spectroscopic observations.

### 2.2 Incompleteness

Incompleteness can arise in several ways and, if it were systematic with redshift or spectral type, might seriously affect LF estimation. The most benign effect, which can be corrected, is incompleteness that arises purely from the increased difficulty of making redshift identifications because the spectra of the fainter galaxies in each of the various magnitude ranges have inadequate signal/noise. *Provided* this magnitude-dependent incompleteness is independent of redshift or type, then it can be corrected by weighting each galaxy inversely with the survey success rate at that apparent magnitude. The completeness as a function of apparent magnitude for the various surveys is shown in Figure 1. All the surveys show some drop in completeness at the faint end of their magnitude range. The worst-affected surveys are AF-bright and LDSS-2, while DARS is virtually complete. The relatively low completeness of the AF-bright survey arises from our strategy of doing the observations for this survey whenever the conditions were too poor for the AF-faint survey. As a consequence, the AF-bright spectra are often of poorer quality than the AF-faint spectra.

We can estimate the effect of the observed incompleteness (and the efficacy of a magnitude-dependent completeness correction of the type described above) by comparing the distributions of the $V/V_{max}$ statistic for the various data subsets with and without the correction for magnitude-dependent incompleteness. If the observed distribution of galaxies is unclustered and does not evolve then $V/V_{max}$ should be uniformly distributed between 0 and 1. Actual clustering and evolution will cause departures from this expectation, but so will magnitude-dependent incompleteness even in their absence.

The form of departure from uniformity of the $V/V_{max}$ distribution is different for each of these cases. Magnitude-dependent incompleteness will cause the sample to be deficient in the higher redshift galaxies of any given luminosity, and will therefore bias the $V/V_{max}$ distribution to smaller values; clustering will cause peaks and troughs in the distribution at the values of $V/V_{max}$ corresponding roughly to an $L^*$ galaxy at the redshift of the relevant structure; evolution (at least if it takes the form of an increase in the number of galaxies of any given luminosity at higher redshifts) will bias the distribution to larger values. Note that an important feature of our strategy of breaking our samples into several narrow apparent magnitude slices, is that we expect little relative evolution over any one subsample. Only by combining all the surveys and spanning a large range in apparent magnitude and redshift do we expect to see evidence for evolution. Thus the absence of any upward trend within a survey covering a narrow apparent magnitude range is not evidence against evolution.

**Table 1.** The redshift surveys.

| Survey | $b_J$ | Area $\square^\circ$ | Fields | Gals | ID% | $\langle V/V_{max}\rangle$ raw | corr | $d\langle V/V_{max}\rangle$ | $n$ | $m$ | $n'$ |
|---|---|---|---|---|---|---|---|---|---|---|---|
| DARS     | 11.5–17.0 | 70.840 | 5  | 328 | 96% | 0.46 | 0.46 | 0.016 | 2.5 | 3.5 | 1.3 |
| AF-bright| 17.0–20.0 | 5.519  | 16 | 478 | 70% | 0.43 | 0.48 | 0.013 | 1.8 | 1.9 | 1.3 |
| AF-faint | 19.5–22.0 | 4.670  | 16 | 548 | 81% | 0.45 | 0.46 | 0.012 | 3.6 | 1.5 | 2.9 |
| BES      | 20.0–21.5 | 0.499  | 5  | 188 | 83% | 0.44 | 0.47 | 0.021 | 1.4 | 0.8 | 1.4 |
| LDSS-1   | 21.0–22.5 | 0.124  | 6  | 100 | 82% | 0.44 | 0.46 | 0.029 | 1.4 | 1.3 | 1.2 |
| LDSS-2   | 22.5–24.0 | 0.096  | 7  | 84  | 72% | 0.48 | 0.52 | 0.038 | 0.5 | 1.6 | 0.4 |

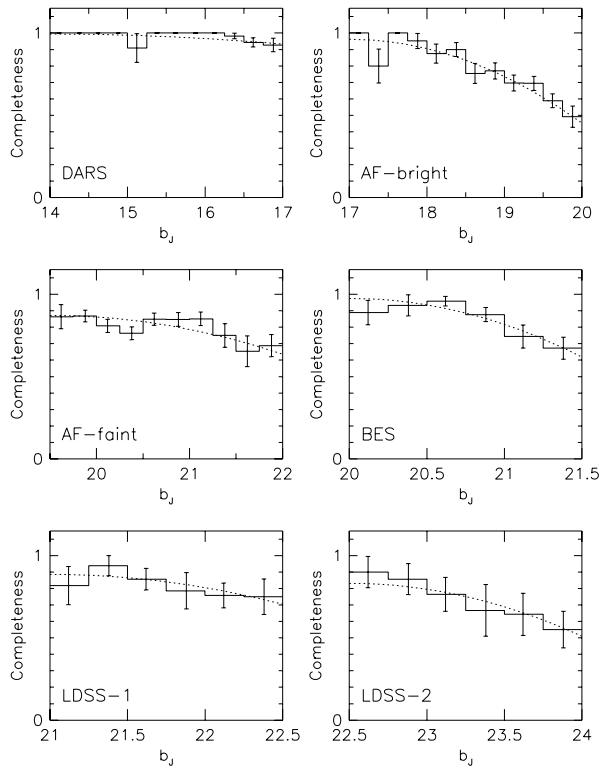

**Figure 1.** Completeness as a function of apparent magnitude for the various surveys. The dotted line is the fit used in applying the magnitude-dependent completeness correction.

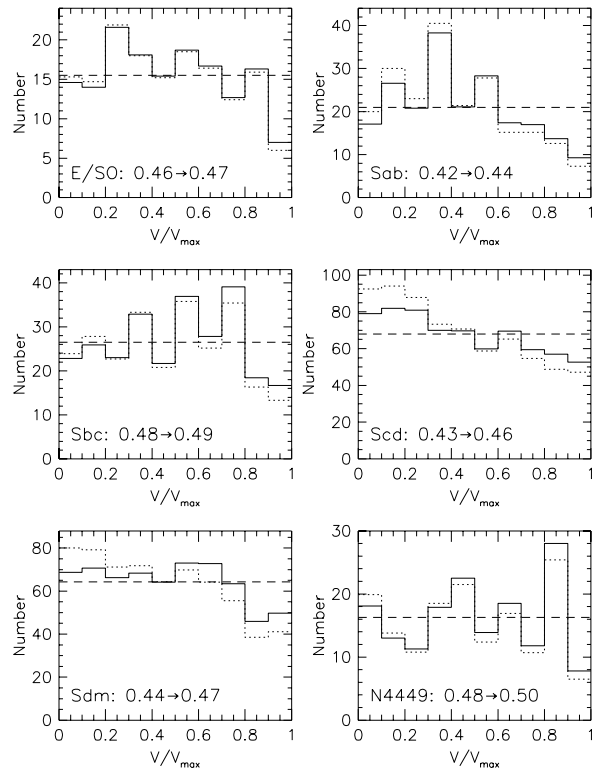

**Figure 2.** $V/V_{max}$ distributions for each spectral type. The dotted lines show the distributions before applying the magnitude-dependent completeness corrections and the solid lines after. The values of $\langle V/V_{max}\rangle$ before and after the correction are indicated.

Table 1 lists the mean value of the $V/V_{max}$ statistic for each survey before and after applying the correction for magnitude-dependent incompleteness (which is shown as the dotted lines in Figure 1). Uncertainties refer to standard errors in the mean of $N$ instances of a uniform random variable, viz. $\sqrt{\frac{1}{12N}}$. The table indicates the significance with which our observed values (after correction) depart from the expectation value of 0.5.

Clustering increases the uncertainty of this test. If there are typically $m$ objects per cluster, the uncertainty in $V/V_{max}$ becomes $\sqrt{\frac{m}{12N}}$. We can estimate $m$ very crudely by considering the observed standard deviations $s$ in the $V/V_{max}$ histograms which, for 10 bins, becomes $m = 10s^2/N$. With these values of $m$, the revised $n' = \frac{n}{\sqrt{m}}$ is consistently less than 3, suggesting no significant non-uniformities remain in the completeness-corrected samples. We will later demonstrate that the effect of this remaining incompleteness on the LFs is small.

Unlike magnitude-dependent effects, incompleteness that is a function of galaxy redshift or spectral type can neither be directly quantified nor corrected. Furthermore both these forms of incompleteness may be confused with the signal/noise-dependent losses, since both type and redshift are expected to correlate with apparent magnitude. However we can make tests to establish whether either of these problems is significant.

For type-dependent incompleteness we can again use the $V/V_{max}$ statistic. In Section 3.1 we define a procedure to allocate a spectral type to each galaxy by correlating its spectrum with local templates. Anticipating this classification scheme, Figure 2 shows $V/V_{max}$ distributions for each spectral type (as defined in Section 3.1) with and without the correction for magnitude-dependent incompleteness. In every case the correction leads to more uniform $V/V_{max}$ distributions (i.e. $\langle V/V_{max} \rangle$ closer to 0.5), although a slight deficit of objects with large values of $V/V_{max}$ still remains.

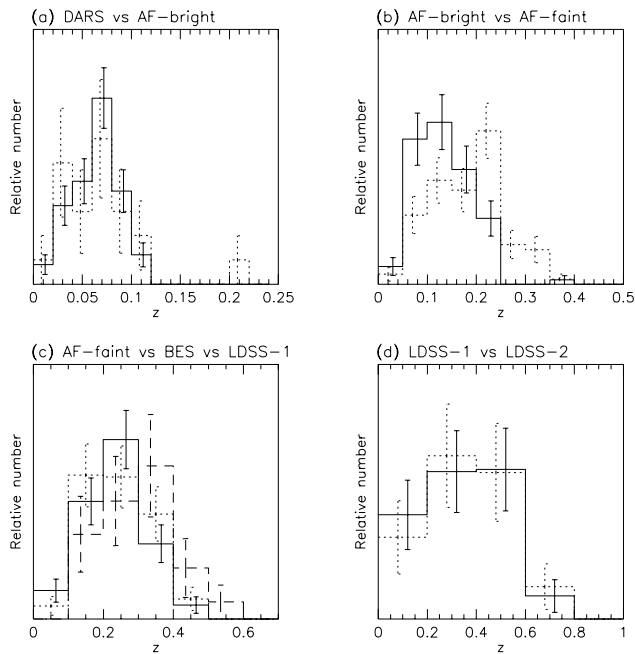

**Figure 3.** Comparison of the redshift distributions in the overlap magnitude ranges of the various surveys. In each panel the first survey is the solid line, the second dotted and the third dashed. The distributions are normalized to have the same total number of objects. Poisson error bars are shown. (a) DARS $b_J=16.5-17$ versus AF-bright $b_J=17-17.5$; (b) AF-bright versus AF-faint, both in $b_J=19.5-20.0$; (c) AF-faint versus BES versus LDSS-1, all with $b_J=21-21.5$; (d) LDSS-1 $b_J=22-22.5$ versus LDSS-2 $b_J=22.5-23$.

For redshift-dependent incompleteness the $V/V_{max}$ statistic is inapplicable because $V$ is a function of $z$. However we can check for redshift-dependent incompleteness by making use of the important fact that our combined sample is made up of sub-surveys with overlapping apparent magnitude ranges. By comparing the redshift distribution of the bright (high-completeness) end of a fainter survey with the faint (low-completeness) end of a brighter survey we can, within the limits imposed by clustering, check whether incompleteness distorts the redshift distributions. By restricting the LF analyses to those based on data within limited redshift ranges, we can further limit the effect of such incompleteness.

Figure 3 shows the results of such comparisons. With the exception of the overlap between AF-bright and AF-faint there is good agreement between the redshift distributions, implying that redshift-dependent incompleteness is not a problem. Of course we cannot check the LDSS-2 survey in this way since we have no fainter survey with which to compare it. Glazebrook et al. (1995a) discuss the limitations of this deepest data set in some detail. The significant difference between the AF-bright and AF-faint data in the range $b_J=19.5-20$ is difficult to understand. It seems difficult to attribute this to redshift incompleteness given the ranges involved ($z \simeq 0.1$ in AF-bright c.f. 0.2-0.3 in AF-faint). Conceivably this is a clustering effect or arises from the small sample sizes.

To summarise, there is significant incompleteness in all the surveys included in this work. However this incompleteness appears to be dominated by the difficulty of identifying the fainter galaxies in each sample due to poorer spectral S/N. We can remove this effect satisfactorily by applying a magnitude-dependent completeness correction. Although some residual systematic effects remain, these are small; we later show that even the dominant magnitude-dependent correction does not seriously affect our LF results.

## 3 ANALYSIS

The full Autofib survey catalogue containing positions, photometry and spectral classifications will be published in Paper III. The raw data for analysis consists of galaxy positions precise to better than 0.5 arcsec rms, $b_J$ magnitudes and redshifts. The first and most important step in determining the galaxy LF is calculating the luminosity. Once a cosmological framework has been selected (we adopt $q_0=0.5$ and $H_0=100h$ km s$^{-1}$ Mpc$^{-1}$), the distance modulus for each galaxy can be readily determined. However, in samples at moderate redshift, the $k$-correction is a very significant term and a strong function of spectral class and redshift. For the range of Hubble types seen locally, the $k$-correction for the $b_J$ system ranges from 0-2 mag at the mean redshift of the LDSS-2 data, and 0-1 mag even at the mean redshift of the AF-faint data. In order to make progress, therefore, we also need to define a robust classification procedure from which type-dependent $k$-corrections can be estimated for every galaxy in the survey.

### 3.1 $k$-corrections

Previous researchers have used a variety of approaches to estimate $k$-corrections. The most common method is to assume that galaxies have $k$-corrections that increase linearly with redshift, with each morphological type assigned a different slope (e.g. Efstathiou et al. 1988 and Loveday et al. 1992). If colours are available, the observed colour and redshift can be used to infer the spectral type by comparison with predictions from a set of template spectral energy distributions, and the $k$-correction then follows (e.g. Colless et al. 1990). In a precursor analysis to that carried out here, Eales (1993) used the alternative approach of calculating luminosities in a passband corresponding to the $b_J$ band shifted blueward by the mean redshift of the sample. This has the advantage that errors in the $k$-correction are minimised as the correction at the mean redshift is defined to be zero. However, Eales was unable to assign types for any but the nearest galaxies in his analysis (those in DARS) and thus his luminosities could be in error by as much as 1 mag.

For the Autofib redshift survey the above-mentioned methods for obtaining $k$-corrections are either inapplicable or inadequate. Only the DARS galaxies are bright enough for morphological classification and only the LDSS-1 and LDSS-2 samples have $b_J-r_F$ colours, while applying a mean $k$-correction or using a mean redshift gives large errors in the inferred luminosities and even larger errors in the volume weighting necessary to recover the LF.

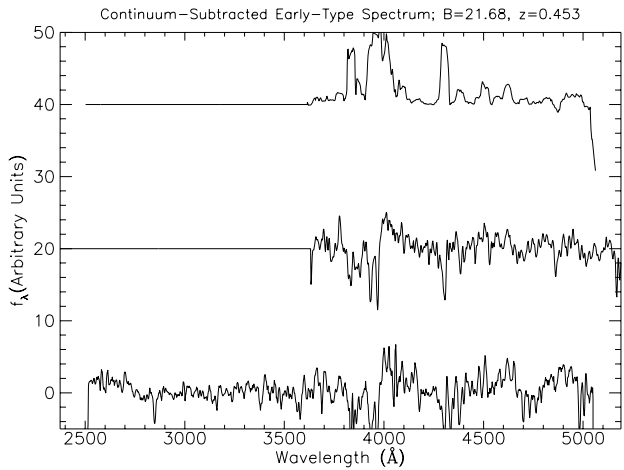

**Figure 4.** Example of spectral classification by the cross-correlation method. The middle curve shows the continuum-subtracted spectrum of a $z=0.453$ galaxy with $b_J=21.68$ from the AF-faint survey. The lower curve is the best-matching template spectrum from the Kennicutt atlas, which belongs to an early-type (Sab) galaxy. The upper curve shows the cross-correlation of the two spectra.

The ideal solution would be to derive the $k$-correction directly from each spectrum. To do this we would need to sample the $b_J$ response curve ($\lambda\lambda$ 3800–5400 Å ) in both the observed and rest frames. However for high-redshift objects the restframe $b_J$ lies outside our spectral range. We would also need to have reliably flux-calibrated data, but this is difficult at faint limits, where the sky-subtraction introduces uncertainties that make the spectra adequate only for identifying features.

Clearly the way forward is to *classify* the spectra and relate this classification to a well-defined set of $k$-corrections. Rather than relying on specific spectral features (which may not always be present), we chose to cross-correlate the survey spectra against those of the Kennicut (1992a, 1992b) spectral library of similar spectral resolution. These library spectra are well-suited for use as cross-correlation templates because their wavelength coverage is well-matched to our survey spectra and because they sample the integrated light of the galaxies, which is approximately also the case for our fibre and slit spectra of faint galaxies.

Prior to cross-correlation, the Kennicut template spectrum and the survey spectrum were smoothed on a 100Å scale in the observer's frame. The smoothed versions were then subtracted away, yielding continuum-subtracted spectra rebinned to 2Å per pixel. The survey spectrum was then assigned the type of the template with which it most strongly cross-correlates. The published morphology of the appropriate Kennicut template indicates which of the King & Ellis (1985) $k$-corrections is used for that particular survey spectrum. This table of $k$-corrections is available for E/S0, Sab, Sbc, Scd, Sdm types and for NGC4449, the latter being an intense star-forming galaxy representative of the bluest classes identified in our survey. An illustration of this method is given in Figure 4.

To check this algorithm, we performed a series of simulations. A Kennicut spectrum was selected at random and normalised to a suitable mean count per pixel. This spectrum was next redshifted by a random $z$ between 0 and 0.6, multiplied by an approximation to the instrumental response function and then brought back to zero redshift. Finally, the observed spectrum was generated as a set of random Gaussian deviates about this modified template spectrum with a S/N per pixel in the range 0.8–4.0. These test spectra were processed similarly to the real survey spectra. The success rate in identifying the correct spectral type was highly satisfactory: averaging over all redshifts, the success rate was 70% for spectra with S/N=1 per pixel and >80% for spectra with S/N>2 per pixel; averaging over all S/N levels, the success rate is >80% for $z<0.5$; for $z>0.5$ the success rate drops to 40%, however, a consequence of the lack of overlap in the restframe between the templates and the observed spectra.

We therefore classified the galaxy spectra from the various surveys as follows: (i) for DARS we used the morphological types given by Peterson et al. (1985); (ii) for AF-bright, AF-faint and BES we used the cross-correlation method described above; (iii) for LDSS-1 we used the cross-correlation method supplemented by the use of the published $b_J-r_F$ for galaxies that were either at too high a redshift or had too low a S/N for the method to be reliable; (iv) for LDSS-2 we used the published $B-R$ colours to infer spectral types. For the 136 galaxies where we could not classify a spectrum with the cross-correlation method and did not have either a morphological type or colour, then we used the $k$-correction appropriate to an Scd (the median spectral type of the whole survey) in computing its luminosity.

As an external check on the cross-correlation classifications, we can compare the $b_J-r_F$ colour observed for those galaxies in the LDSS-1 survey (Colless et al. 1990) with the colour predicted from the galaxy's redshift and its spectral type as derived by the cross-correlation method (see Figure 5). The agreement is generally very good: the rms scatter of 0.4 mag reflects both the expected 0.2 mag rms uncertainties in the observed colours and a small number of objects with odd colours resulting from image mergers on the plate, as well as the errors in the spectral classifications.

A detailed description of the spectral classification algorithm and more exhaustive tests of the method are given in Paper II. However, we can illustrate the precision attained by assuming 20% of the galaxies are misclassified by one class equally in both directions - an error consistent with the discussion in Paper II. We can then calculate the rms $k$-correction error for a given redshift bin and class from the differential trends with class, including an allowance for the fact that the class is a discrete approximation to the actual spectral energy distribution. The errors are weighted by the numbers in each class to give the rms error plotted in Figure 6. This error increases with $z$ but is comparable to the photometric errors over the redshift range of the samples.

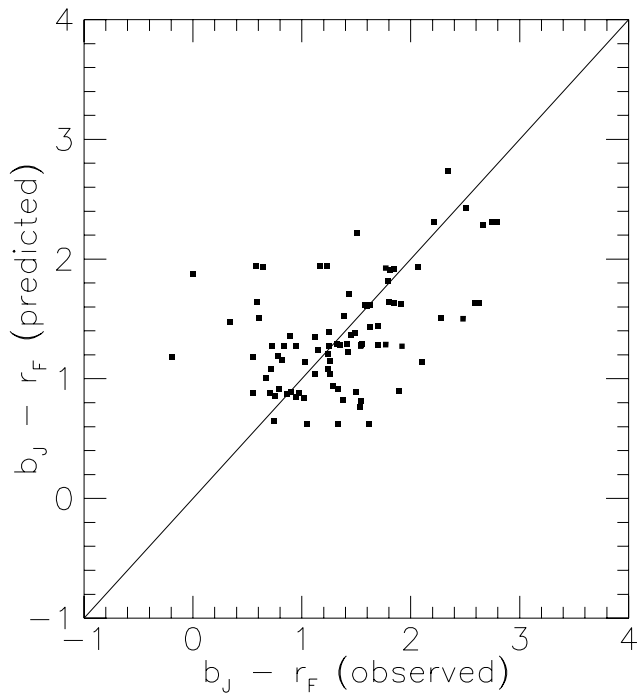

**Figure 5.** Comparison of the observed $b_J - r_F$ colours of LDSS-1 survey galaxies with their colours as predicted from their redshifts and cross-correlation spectral types.

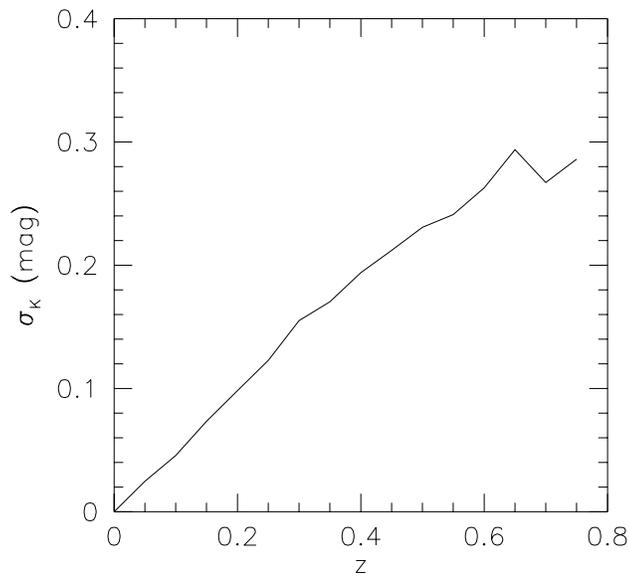

**Figure 6.** The rms error in the $k$-correction as a function of redshift assuming 20% of the galaxies are misclassed by $\pm 1$ spectral class (see text for details).

### 3.2 Luminosity Function Estimation

In our analyses we have used two related methods to estimate luminosity functions: the traditional $1/V_{max}$ method and a modified version of the step-wise maximum likelihood method (SWML). The latter method was developed specifically for our survey and fulfills our requirement for extracting the luminosity function within various redshift ranges from a number of catalogues lying within various magnitude limits. The $1/V_{max}$ method is more direct and provides an unbiased maximum-likelihood/minimum-variance estimate of the luminosity function in the absence of clustering; SWML allows one to trade resolution in absolute magnitude for insensitivity to clustering. In the limit of small magnitude bins the two methods become identical. A full description of the modified SWML method, and a comparison of $1/V_{max}$ and SWML with other techniques for estimating luminosity functions, are given in Paper II. Here, for simplicity, we present results based on the $1/V_{max}$ method. None of the conclusions in this paper are sensitive to the LF estimator used.

The $1/V_{max}$ method is the canonical direct estimator of the luminosity function, first introduced by Schmidt (1968) for the study of quasar evolution (see also Felten 1976). Avni & Bahcall (1980) showed how to combine more than one sample in a $1/V_{max}$ analysis, and Eales (1993) extended the method to construct the luminosity function as a function of redshift. The method works as follows.

Suppose we have $N$ galaxies and for each galaxy $i$ we have measured its apparent magnitude $m_i$ and its redshift $z_i$. These galaxies were obtained in $M$ samples, and sample $j$ covers an apparent magnitude range $m_{1j} \leq m \leq m_{2j}$ and an area (solid angle) of sky $\omega_j$ (in steradians). It also has a sampling rate $S_j$ (the fraction of galaxies in the given magnitude range and area that were observed) and a completeness $C_j$ (the fraction of the observed galaxies for which redshifts were obtained). Any *known* dependence of the sampling rate or completeness on apparent magnitude, redshift or spectral type can be removed by appropriate weighting.

The luminosity function (number of galaxies per unit comoving volume per unit magnitude) in the absolute magnitude range $M_1 \leq M \leq M_2$ and redshift range $z_1 \leq z \leq z_2$ can then be estimated as

$$\frac{\int_{M_1}^{M_2} \int_{z_1}^{z_2} \phi(M,z)\,dz\,dM}{(M_2 - M_1)(z_2 - z_1)} = (M_2 - M_1)^{-1} \sum_{\{i: M_1 \leq M_i \leq M_2\}} 1/V_i \quad (1)$$

where the sum is over galaxies in the given absolute magnitude range and $V_i$ is the total accessible volume of galaxy $i$. This volume is

$$V_i = \sum_{j=1}^{M} V_{ij} \quad (2)$$

where

$$V_{ij} = \Omega_j \int_{z_{min}^{ij}}^{z_{max}^{ij}} \frac{dV}{dz}\,dz \quad (3)$$

is the accessible volume of the galaxy $i$ in sample $j$ and $\Omega_j = \omega_j S_j C_j$ is the effective area in steradians of this sample. In this way we treat the $M$ samples as a single coherent sample (following Avni & Bahcall 1980). The integral is over the comoving volume element (see below) and the limits are the lowest and highest redshifts at which galaxy $i$ remains both within sample $j$'s magnitude range $m_{1j} \leq m \leq m_{2j}$ and within the redshift range $z_1 \leq z \leq z_2$. If $z(M, c, m)$ is the redshift at which a galaxy of absolute magnitude $M$ and spectral class $c$ has an apparent magnitude $m$, then

$$z_{min}^{ij} = \max[z_1, z(M_i, c_i, m_{1j})] \qquad (4)$$

and

$$z_{max}^{ij} = \min[z_2, z(M_i, c_i, m_{2j})]. \qquad (5)$$

For completeness, we note that the absolute and apparent magnitudes of galaxy $i$ are related by

$$M_i = m_i - 5 \log d_L(z) - K(z, c_i) - A_i - 25 \qquad (6)$$

where $A_i$ is the Galactic absorption in the direction of the galaxy (which we assume to be negligible throughout our analysis), $K(z, c_i)$ is its $k$-correction and $d_L(z)$ is its luminosity distance in Mpc, given by

$$d_L(z) = \frac{cz}{H_0} \left[ \frac{1 + z + (1 + 2q_0 z)^{1/2}}{1 + q_0 z + (1 + 2q_0 z)^{1/2}} \right]. \qquad (7)$$

The volume element (in Mpc$^3$) corresponding to a solid angle of 1 steradian and a thickness of $dz$ at redshift $z$ is

$$\frac{dV}{dz} = \frac{c}{H_0} \frac{d_L^2}{(1 + z)^3 (1 + 2q_0 z)^{1/2}}. \qquad (8)$$

As shown by Felten (1976), the $1/V_{max}$ method is an unbiased, maximum-likelihood, minimum-variance estimator of the luminosity function. However clustering in the galaxy sample causes the $1/V_{max}$ estimator to produce spurious 'features' due to the assumption that the galaxy number density is everywhere constant (apart from a possible evolutionary variation with redshift). Thus a cluster at low redshift will be misinterpreted as an excess of intrinsically faint galaxies, while a cluster at high redshift will produce a spurious excess of luminous galaxies.

The uncertainties in the luminosity functions derived by the $1/V_{max}$ method can be obtained either using the approximate formula given by Felten (1976) or (as we have done here) by using standard bootstrap error estimation techniques. Note that we have not applied any corrections to our LFs for the photometric errors in our magnitudes. This is because (i) these corrections would be small, since the rms photometric errors are typically 0.1–0.2 mag, which is much smaller than the 0.5 mag bins we use for computing the LFs, and (ii) because uncertainties in the $k$-corrections are at least as large. We consider the effects of the latter in more detail below.

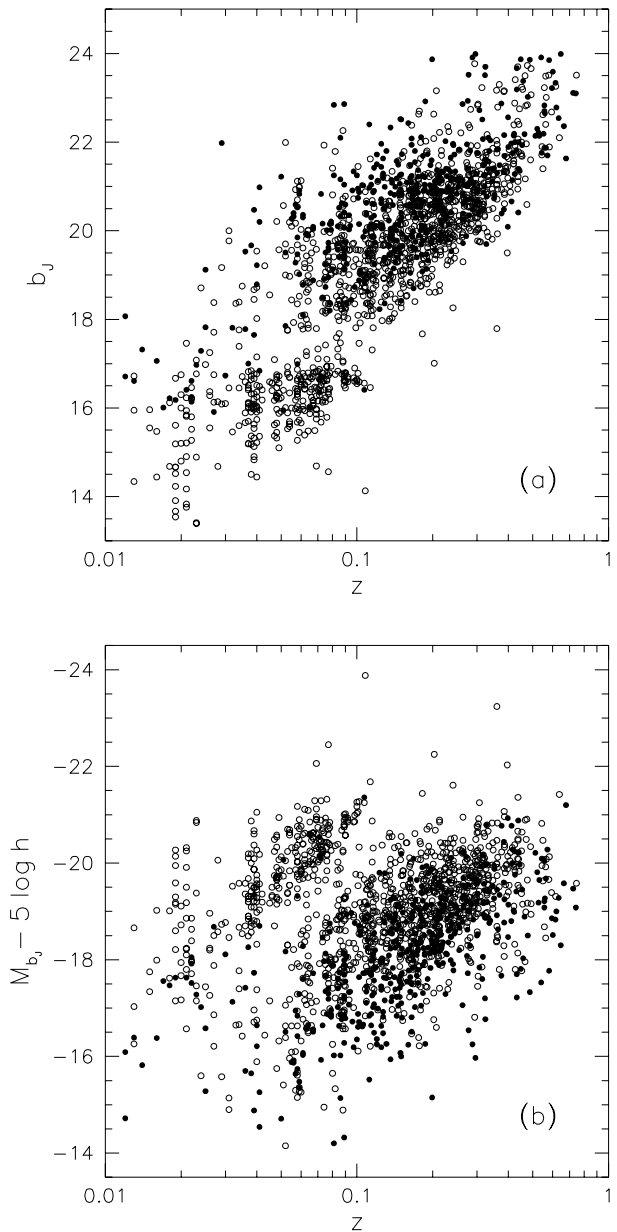

**Figure 7.** The survey data: (a) apparent magnitude–redshift distribution, and (b) absolute magnitude–redshift distribution. Galaxies with strong [OII] emission (those with rest-frame equivalent widths $W_\lambda \geq 20$Å) are shown as filled circles.

## 4   RESULTS

The distribution of absolute magnitude with redshift for the entire survey is shown in Figure 7. Although it is not straightforward to interpret because of the various samplings, solid angles and magnitude limits of each sub-survey and the effects of $k$-corrections on the relative numbers of different galaxy types, two important results (which we establish rigorously below) are already apparent. Firstly, there appears to be a dearth of sources at the faint end of the LF locally, notwithstanding the very faint apparent magnitude limits now probed by LDSS-1 and LDSS-2. This suggests there is no significant population of low-luminosity sources (see the discussion by Glazebrook et al. 1995a). Secondly, considering those sources with strong [OII] emission, we note that both the abundance and mean luminosity of these star-forming galaxies appear to increase with redshift. In this section we examine what our combined survey can tell us about (i) the local luminosity function, (ii) the evolution of the luminosity function with redshift, and (iii) the relative evolution of the star-forming galaxies compared to the entire sample.

### 4.1 The Local Luminosity Function

There has been considerable debate on the question of whether the faint end slope of the local luminosity function has been underestimated. The motivation arises partly from theoretical expectations based in hierarchical cosmologies where the required growth of structure can be seeded by dark matter halos but only with an associated steep mass spectrum, corresponding to $\alpha \approx -1.3$ to $-1.5$ (Kauffmann et al. 1994). Elaborate mechanisms are required to circumvent this problem (Cen & Ostriker 1994).

Recent LF estimates (Efstathiou et al. 1988, Loveday et al. 1992, Marzke et al. 1994) consistently indicate a Schechter slope for all galaxies of $\alpha \simeq -1.1$ down to $M_B = -17 + 5\log h$. This is in marked contrast to the LF computed for the nearby Virgo cluster (Binggeli et al. 1988), and so the question has been raised as to whether the local field LF determinations have missed an abundant population of low luminosity sources (cf. McGaugh 1994 and references therein). In their analysis of the CfA redshift survey, Marzke et al. (1994) claim the first evidence for a possible upturn fainter than $M_{Zwicky} = -16 + 5\log h$. Specifically, they observe 3 times as many low luminosity objects in this category as would be expected from an extrapolation of the $\alpha = -1.1$ Schechter function fitted at higher luminosities. We consider the uncertainties are still too great for Marzke et al.'s result to be considered definitive. A scale error in the photometric scale of the Zwicky catalogue could significantly reduce the excess and the volume sampled at these absolute magnitudes is very small. Indeed, the effect is greatest in the northern cap where Virgo galaxies inevitably contaminate the supposed field sample.

It is important here to distinguish between two distinct uncertainties in the local LF whose effects are often confused. Firstly, as described above, the faint end slope remains uncertain and a steep slope cannot formally be excluded from current data fainter than $M_B = -16$. This uncertainty is largely a consequence of the small volumes probed for galaxies with $M_B > -16$ by all extant surveys. In the combined CfA redshift survey of 10620 galaxies over 2.8 steradians in both hemispheres to $m_{Zwicky} = 15.5$ although 293 galaxies were found with $M_{Zwicky} > -16$, they sample a volume contained within only 10-20 Mpc, which is unlikely to be representative. In the deeper Stromlo-APM 1:20 survey of 1769 galaxies over 4300 deg$^2$ to $b_J = 17.15$, the depth is clearly greater, but the number fainter than $M_B \simeq -16$ is only 49. Notwithstanding these uncertainties, a steeper local LF would greatly increase the observed number of apparently faint galaxies, as intrinsically faint sources contribute a Euclidean number count slope (Kron 1980, Phillips & Driver 1995). However, as discussed by Broadhurst et al. (1988), a very significant contribution of low luminosity galaxies at $b_J > 21$ would distort the field redshift distribution to lower values than expected. Clearly only more extensive surveys beyond $B = 17$, such as that discussed here, can resolve this issue definitively.

A second, and independent, uncertainty has been pointed out by many workers (e.g. Ferguson & McGaugh 1995), namely that many field surveys may miss altogether a population of low surface brightness galaxies (LSBGs) by virtue of selection effects inherent in standard image detection algorithms (Disney & Phillipps 1985, Davies et al. 1989). As an undetected population, their location in the LF is a matter of conjecture. Most direct searches for LSBGs have found relatively few compared to the numbers of galaxies of normal surface brightness (Dalcanton 1994, Roukema & Peterson 1994). However Schwartzenberg et al. (1995) have recently claimed to find ten times as many LSBGs as normal galaxies brighter than $0.1L^*$. This claim requires further investigation, as it depends critically on indirect estimates for the redshifts of the objects involved.

Although one might assume that LSBGs might lie predominantly at the faint end of the LF, thereby linking with the problem discussed above, this need not necessarily be the case. Indeed, some of the LSBGs so far identified are fairly luminous (Bothun et al. 1989). If the surveys conducted at faint apparent magnitudes systematically probed to lower surface brightness limits, they might reveal a higher volume density of galaxies over a range of luminosities, and hence more faint galaxies. In the rather unlikely case of similar LFs for the high and low surface brightness populations, the hypothesis could be tested with surface brightness profiles at various redshifts and magnitudes. Broadly speaking, one would expect to uncover more LSBGs at fainter limits.

The most straightforward argument *against* the faint galaxy population being dominated by LSBGs comes from recent ground- and space-based observations with sufficient resolution to determine the sizes of faint galaxies. Colless et al. (1994) found that the size-luminosity relation for a sample of 26 $b_J \approx 22$ galaxies drawn from the LDSS-1 survey, with redshifts up to $z \approx 0.7$, was entirely consistent with that of normal low-redshift spirals. Likewise, preliminary HST studies of galaxies to $I \approx 21$ (Mutz et al. 1994, Phillips et al. 1995) also show a stable size-luminosity relation and no excess of LSBGs.

Figure 8 shows the local ($z < 0.1$) LFs derived from the DARS survey and from the combined AF-bright and AF-faint surveys (hereafter Autofib). The solid and short-dash curves are the Schechter function fits to the Stromlo-APM survey by Loveday et al. (1992) and to the DARS survey by Efstathiou et al. (1988) respectively. The parameters of these fits are $M_{b_J} = -19.50$, $\alpha = -0.97$, $\phi^* = 0.014\,h^3\,\mathrm{Mpc}^{-3}$ for Stromlo-APM and $M_{b_J} = -19.56$, $\alpha = -1.04$, $\phi^* = 0.008\,h^3\,\mathrm{Mpc}^{-3}$ for DARS. The fits apply to the range $-22 \leq M_{b_J} \leq -17$; the dotted curves show the extrapolations to fainter magnitudes. The long-dashed curve is the Schechter function fit to the Autofib survey LF over the range $-20 \leq M_{b_J} \leq -14.5$, and has parameters $M_{b_J} = -19.20^{+0.29}_{-0.34}$, $\alpha = -1.09^{+0.10}_{-0.09}$, $\phi^* = 0.026^{+0.08}_{-0.08}\,h^3\,\mathrm{Mpc}^{-3}$ ($\chi^2 = 11.6$ for 10 degrees of freedom).

The DARS and Stromlo-APM LFs agree at the bright

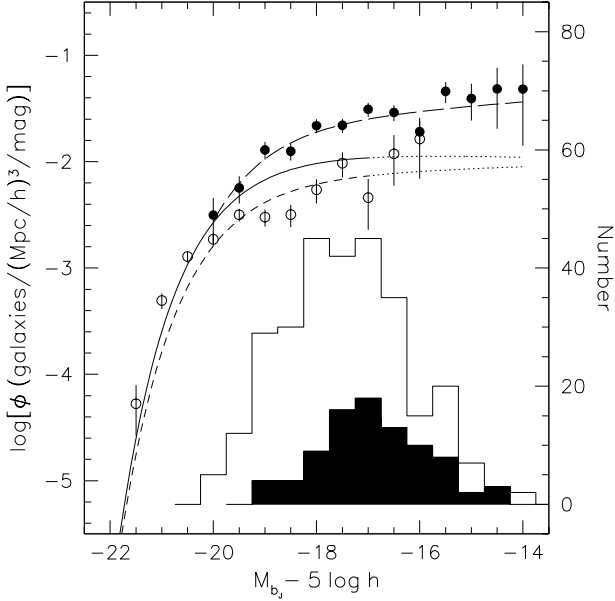

**Figure 8.** The local ($z<0.1$) luminosity functions from the DARS survey (open circles) and combined surveys excluding DARS (filled circles). The solid curve is the Loveday et al. fit to the Stromlo-APM survey LF and the short-dash curve is the Efstathiou et al. fit to the DARS LF (the dotted curves are the extrapolations of these fits from $M_{b_J}=-17$ to $-14.5$). The fit to the combined LF (excluding DARS) is shown as the long-dash curve; the open histogram shows the absolute magnitudes of the galaxies contributing to this LF, while the shaded histogram shows the distribution of galaxies with $W_\lambda[{\rm OII}]>20$Å.

and faint ends, but DARS has a deficit of galaxies with $-20 < M_{b_J} < -18$. This deficit leads to the 2× lower normalization and mildly steeper faint-end slope in the Schechter fit to DARS. The low normalization of the DARS LF was noted by Efstathiou et al. in comparing the DARS counts with those of deeper photometric surveys; they claimed it was marginally consistent with the effects of clustering if uncertainties in the selection function were also taken into account.

The Autofib $z<0.1$ LF is signicantly higher than the Stromlo-APM LF everywhere fainter than $M_{b_J}=-19.5$; the LF is not well-defined brighter than this, where it is determined from only 17 galaxies. The faint end is again flat at least as faint as $M_{b_J}=-16$, but has a normalization that is about a factor 2.5 higher than Stromlo-APM or DARS.

How can one interpret this change in the normalization between the DARS/Stromlo-APM LF and the Autofib LF? If it is due to evolution it is remarkably rapid: the galaxies at, say, $M_{b_J}=-17$ in the $b_J<17$ surveys are at $z<0.02$ while galaxies of the same luminosity in the Autofib surveys are close to the redshift limit imposed on this 'local' LF (i.e. $z=0.1$), corresponding to a lookback time of only $0.9\,h^{-1}$ Gyr. An alternative explanation is some sort of measurement error in the bright or faint survey magnitudes, including residual isophotal effects associated with the fainter surface brightness thresholds associated with the deeper survey data as advocated by Metcalfe et al. (1995b). This explanation also poses difficulties since a zeropoint offset between the various surveys produces a horizontal rather than vertical shift in the LFs, while a magnitude scale error would produce a change in the faint-end slope.

The higher normalization in the observed local LF for surveys at fainter magnitude limits is the direct counterpart to the well-known observation that the number counts at $b_J<19$ are much steeper than is predicted by a model with a non-evolving LF with a flat faint end. The solution to this puzzle is unclear, but one suggested resolution can be ruled out: the steep counts are *not* due to a non-evolving LF with a steep faint end (at least, not down to $M_{b_J}=-16$). Such a conclusion might be incorrectly drawn if one simply combined two surveys with different magnitude limits on the assumption that the LF does not change: combining the DARS and Autofib surveys to produce an overall $z<0.1$ LF results in a misleadingly steep faint-end slope of $\alpha=-1.3$ despite the fact that each survey has $\alpha=-1.0$ because the bright end of such a combined LF is dominated by the low-normalization DARS LF while the faint end is dominated by the high-normalization Autofib LF. These arguments suggest the DARS sample may be unrepresentative and thus we will exclude it in our further analyses.

Evidence for a higher normalisation than that originally suggested by DARS has also come from independent $I$-band redshift surveys (Lilly et al. 1995), morphological-based counts obtained with Hubble Space Telescope (Glazebrook et al. 1995c) and from LFs estimated from galaxies identified on the basis of their MgII absorption lines in distant unrelated QSO spectra (Steidel et al. 1995).

A further significant development from the Autofib survey, however, is that we can comment on the nature of the LF fainter than $M_{b_J}=-16$ more reliably than previous workers. This arises from two specific features of the survey. Firstly, by probing fainter limits we survey deeper and more representative volumes. For example, at $M_{b_J}=-14$, galaxies can be located across all fields of the Autofib survey in a total effective volume of 2600 Mpc$^3$, 3.7 times larger than that appropriate for Marzke et al's CfA survey. Significantly, for the bulk of our survey reaching to $b_J \simeq 22$, such dwarfs would be seen to 160 $h^{-1}$ Mpc (compared to only 8 Mpc in CfA) indicating much more representative volumes when the large number of independent pencil beams spanning the southern sky is taken into consideration (§2.1). Secondly, the faint end of the local LF is probed most effectively from the $b_J>21$ samples which were selected from deep 4-m plates and ancilliary CCD data. This material was thresholded at a low surface brightness limit of approximately $\mu_{b_J}=26.5$ mag arcsec$^{-2}$ (Jones et al. 1991) which would guarantee detection of LSB galaxies of the kind proposed.

Since the samples are still small, the simplest way to proceed is to address the hypothesis that there is an upturn in the LF fainter than $M_B = -16 + 5\log h$, as proposed by Marzke et al. (1994). Figure 9 shows the ratio of the number of galaxies found in the survey at various luminosities (and its formal uncertainty) to that expected for the Schechter function given above (fitted over the brighter luminosity range down to $M_B=-16$). To test the sensitivity to $\alpha$ we have arbitrarily adjusted its value in this comparison to $-1.1$, $-1.3$ and $-1.5$ while keeping $M_B^*$ and $\phi^*$ fixed. We also examined the case where $M_B^*$ and $\phi^*$ are allowed to take their best-fit values. In both cases, we find no evidence for an upturn of the faint end of the LF as claimed by Marzke et al., and a distinctly different behaviour to that identified by Binggeli et al.. The steepest local LF slope consistent with our data has $\alpha \simeq -1.2$.

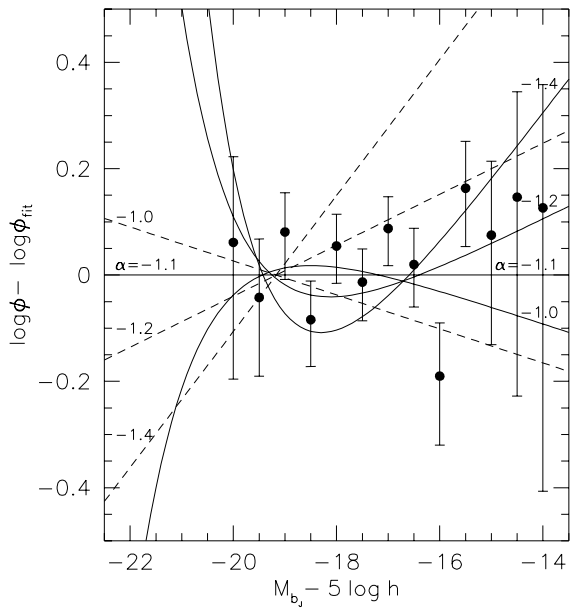

**Figure 9.** The slope of the faint end of the local LF. The points with error bars show the logarithmic difference between the observed local LF from the combined surveys (excluding DARS) and the best-fit Schechter function ($M^*=-19.17$, $\alpha=-1.08$, $\phi^*=0.027$). The various curves are the logarithmic difference between various alternate fits to the LF and this overall best fit. The solid curves are for the best-fitting Schechter functions with $\alpha$ fixed at $-1.0$, $-1.2$ and $-1.4$ (but $M^*$ and $\phi^*$ allowed to vary); the dashed curves are for Schechter functions with the *same* $M^*$ and $\phi^*$ as the overall $\alpha=-1.1$ best-fit, but with $\alpha$ set to $-1.0$, $-1.2$ and $-1.4$.

We also note, in this context, the recent claim for no upturn until at least $M_B > -12 + 5 \log h$ in the Coma cluster (Bernstein et al. 1995), suggesting that the LF of the Virgo cluster as presented by Binggeli et al. may not be representative.

A related question is whether there is any difference between the properties of the intrinsically faint and luminous galaxies, such as might be expected if strong selection effects were limiting the detection of the low luminosity sources. In the *cluster* samples, Binggeli et al. claim the bulk of the low luminosity galaxies are red compact dEs and blue dIrrs. Figure 8 shows that virtually all of the sources fainter than $M_{b_J} \approx -17$ are strong star-forming galaxies with [OII] equivalent widths $W_\lambda > 20$Å. Spectroscopically these are virtually all classified as late-type systems similar to the Virgo dIrrs; no compact red sources are found.

Returning finally to the question of selection biases, it is important to recognise that the surveys most sensitive to the faint end of the local LF are those beyond $B > 21$ including those performed with LDSS-1 and LDSS-2 which address *all* sources, regardless of star/galaxy appearance. Thus compact extragalactic sources would not be missed in these surveys (Colless et al. 1991, 1993).

In summary, we have direct evidence that the absolute scale of the local LF is underestimated by brighter surveys but, significantly, there is *no* evidence for a steeper faint end slope at low redshift in any of the various datasets.

### 4.2 Evolution of the Luminosity Function

Broadhurst et al. (1988) proposed that the redshift distribution of their $b_J$=20–21.5 survey might be reconciled with the excess numbers seen if the luminosity function had a steeper faint end slope in the past (cf. their Figure 9). At what is now a fairly modest magnitude limit, their conclusion was affected by the uncertainty in the absolute normalisation of the LF. An upward shift of a factor of 1.5–2 in $\phi^*$ might remove the need for evolution in the BES data. At the fainter limits probed by LDSS-1 (Colless et al. 1990, 1993) and LDSS-2 (Glazebrook et al. 1995a), the uncertainties in normalisation of the local LF are insufficient to explain the excess counts.

An additional argument used to justify evolution by the above-cited authors was that the *slope* of the counts, $\gamma = d \log N / dm$, is consistently steeper than the no-evolution prediction. Since, in the no-evolution case, $\gamma$ is independent of $\phi^*$, this would appear to provide convincing evidence for some evolution. Unfortunately, this argument fails at some level because no convincing model based on the local LFs has yet reproduced $\gamma$ in the bright, presumably non-evolving, regime $15 < b_J < 20$ where a surprisingly steep count slope is also found (Maddox et al. 1990).

There are two issues that the Autofib survey can address. Firstly, it can be used to directly establish whether there is a change in the LF shape with redshift, and, if so, for which class of sources the evolution is most apparent. Secondly, assuming the fields we have surveyed are representative of those used by Maddox et al. (1990), our survey might cast some light on the question of the true absolute normalisation of the LF, which remains confused.

Figure 10 shows the LFs derived from the $1/V_{max}$ method for three broad redshift bins $0.02 < z < 0.15$, $0.15 < z < 0.35$ and $0.35 < z < 0.75$ corresponding to approximately equal time intervals of about 1.1 Gyr (for $H_0=100$, $q_0=0.5$). (Note that there are only 4 galaxies in the combined sample with $z > 0.75$.) The size and depth of our survey enables us to derive reasonably accurate LFs in all three redshift ranges, although the highest redshift LF only extends to $M_{b_J} \approx -18 + 5 \log h$ whereas the two lower redshift bins extend to at least $M_{b_J} \approx -16 + 5 \log h$. The errors shown were obtained by bootstrap error analysis. In the case of the lowest redshift bin, we have excluded the DARS sample following the discussion in §4.1. The figure clearly shows evidence for a steepening of the faint end slope of the LF with increasing redshift, and perhaps an increase in the overall normalization. However, the trend is not completely clean as the LF at $M_{b_J} \simeq -19$ drops at intermediate redshift and so evidently there are still fluctuations arising from the small sample size.

Formally, 1- and 2-sample $\chi^2$ tests show that, in the region of overlap, the lowest redshift LF in Figure 10 does not differ in shape significantly from the local one (Figure 8) since $P(>\xi^2)=0.85$. The LFs in the higher redshift bins differ from their lower $z$ adjacent bins with $P(>\xi^2)=0.219$ and 0.008 indicating the bulk of the evolution sets in beyond $z \simeq 0.3$. Most of the apparent evolution inferred in earlier work within $0 < z < 0.3$ arose primarily because of the abnormally low LF normalisation. The most significant result arises when we check whether the entire dataset could be consistent with a non-evolving local LF. Considering the $-21.5 < M_{b_J} < -14.5$ LF with $0 < z < 0.75$ in 6 redshift bins, and maintaining a minimum bin count of 5 galaxies per absolute magnitude interval, a $\chi^2$ test rejects a non-evolving LF with a formal probability of $< 10^{-20}$.

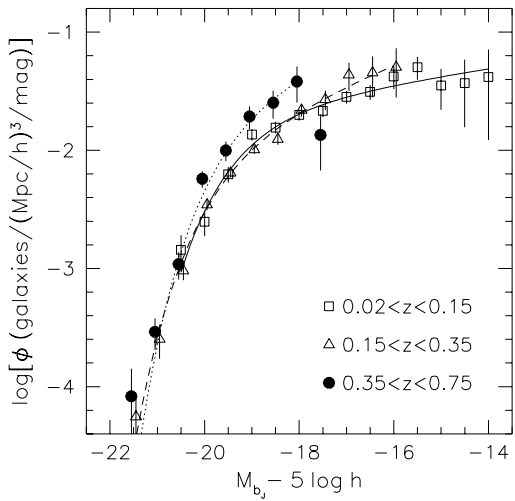

**Figure 10.** Evolution of the luminosity function with redshift. The LF is shown for three redshift ranges corresponding to approximately equal time intervals of about 1.1 Gyr.

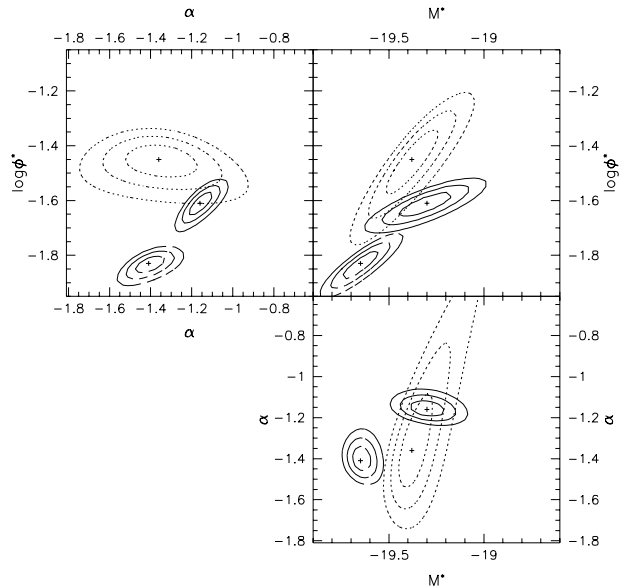

**Figure 11.** Error contours for the pairs of parameters (a) $\alpha$ and $\log \phi^*$, (b) $M^*$ and $\log \phi^*$, and (c) $M^*$ and $\alpha$, fitted to the LFs obtained in the three redshift ranges $0.02 < z < 0.15$ (solid contours), $0.15 < z < 0.35$ (dashed contours) and $0.35 < z < 0.75$ (dotted contours). The contour levels shown are $1\sigma$, $2\sigma$ and $3\sigma$; the crosses indicate the best fits.

The evolution appears to be somewhat stronger for galaxies fainter than $L^*$. The best-fit Schechter functions for each redshift interval are listed in Table 2 and illustrated with their formal error bars in Figure 11. Whereas the trend is not entirely continuous from one redshift range to another (as discussed above), it is important to note that the formal errors do not include any allowance for the possible effects of clustering. Given the small values of $\chi^2/\nu$ and the discussion in §2.2, only a modest correction is expected. Increasing the error bars on Figure 10 by $\sqrt{2}$ would be sufficient to explain the intermediate redshift points at $M_{b_J} \simeq -19$ and would ensure continuity in the Schechter contours of Figure 11. Although a larger sample is ideally required, it is clear from Figures 10 and 11 that the LF has evolved significantly over modest redshifts and it is suggestive that an important component of this evolution is in the faint end slope. There is no convincing evidence for a systematic shift in $M_{b_J}^*$ over $z=0$ to 0.75 whereas $\alpha$ steepens from $-1.1$ to $-1.5$.

Given the potential importance of this result, we need to examine whether it is stable to any procedural uncertainties in our analysis. We have already mentioned that the result is independent of the methods used to compute the LF from our data (see Paper II). In §3.1 we also discuss the small effects that incorrect $k$-corrections (arising from spectral misclassifications) might produce.

The application of the *magnitude*-dependent completeness correction for each survey in fact makes very little difference to the final LFs. The ratio of the LF with the correction to the LF without the correction in each redshift range is shown in Figure 12. The changes are less than about 10% except where the numbers of galaxies contributing to the LF estimate is $\lesssim 10$.

We can place limits on the possible effects of *redshift*-dependent completeness by considering extreme cases where all the unidentified galaxies have either $z=0.05$ or alternatively $z=0.75$. Given our earlier discussion on redshift incompleteness, this must be considered highly unlikely but illustrates the robustness of our main result. If the incompleteness is assumed to be entirely local (Fig. 13(a)), the nearby LF steepens somewhat at the faint end although the evolutionary trends in Fig. 10 are still present. In the case where the incompleteness is assumed to be entirely at high redshift (Fig. 13(b)), an unphysical discontinuity in normalisation with redshift is produced although, again, the evolution seen in Fig. 10 is maintained.

### 4.3 Faint Star-forming Galaxies

Broadhurst et al. (1988, BES) first suggested that the excess population might arise from a distinct population of star-forming galaxies. They noted an increasing number of strong [OII] emission line objects in their survey and claimed these might be sub-$L^*$ galaxies rendered visible during a brief burst of star-formation. By coadding the spectra of several [OII]-strong galaxies, weak Balmer features were identified consistent with this hypothesis. Such a cycle would, however, imply many more quiescent sources were present at fainter magnitudes beyond the limits of current surveys. Conceivably, many galaxies suffered these bursts in the past (with a rate increasing with redshift). Regardless of this, a large population of feeble sources has not yet been seen in local surveys so these galaxies must somehow have disappeared from view. Broadhurst et al. (1992) later demonstrated that the star-forming sources, when separated according to their $W_\lambda$[OII], showed a remarkably steep count slope, whereas the remainder appeared to fit a no-evolution model. This would imply that rapid evolution would predominantly lie in the star-forming galaxies.

**Table 2.** Fitted parameters of the luminosity function with redshift.

| $z$ range | $M^*$ | $\alpha$ | $\log \phi^*$ | $\phi^*$ | $\chi^2$ ($\nu$) |
|---|---|---|---|---|---|
| 0.02<$z$<0.15 | -19.30 [-0.12,+0.15] | -1.16 [-0.05,+0.05] | -1.61 [-0.06,+0.06] | 0.0245 | 8.62 (11) |
| 0.15<$z$<0.35 | -19.65 [-0.10,+0.12] | -1.41 [-0.07,+0.12] | -1.83 [-0.06,+0.08] | 0.0148 | 8.68 ( 9) |
| 0.35<$z$<0.75 | -19.38 [-0.25,+0.27] | -1.45 [-0.18,+0.16] | -1.45 [-0.36,+0.26] | 0.0355 | 4.57 ( 5) |

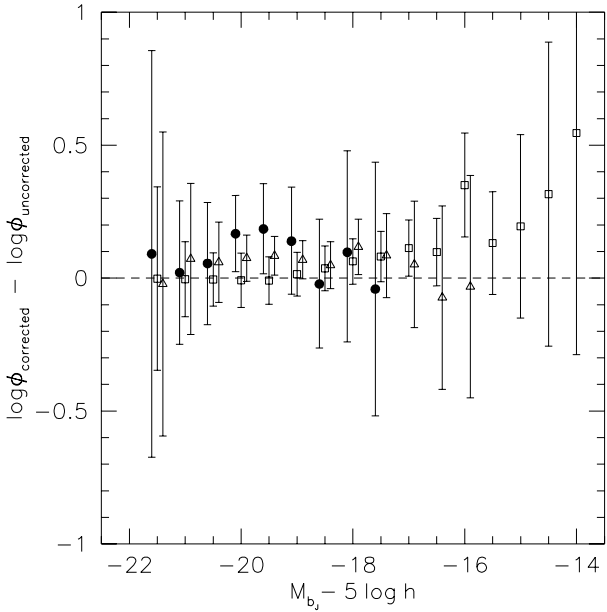

**Figure 12.** Logarithmic difference between the completeness-corrected and uncorrected LFs in the three redshift ranges shown in Fig. 10 (with the same symbols for each range).

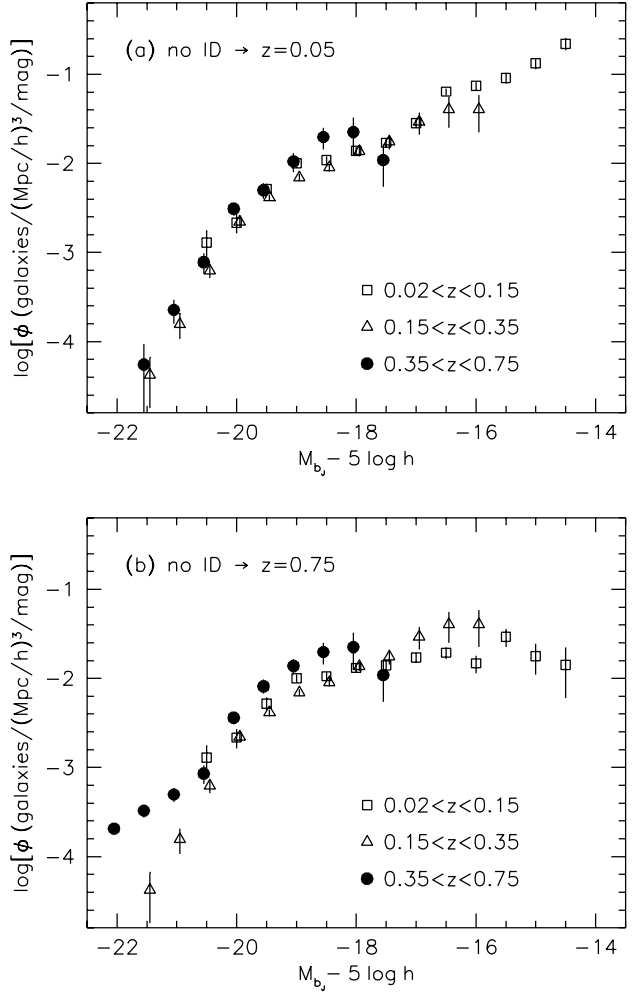

**Figure 13.** The LFs in three redshift ranges as in Fig.10, but assuming that all the unidentified objects have (a) $z$=0.05 or (b) $z$=0.75 and that therefore the sample is 100% complete.

It is important to recognise that discontinuous star-formation events could readily transform a galaxy from being a member of the 'quiescent' population to one with strong [OII] or vice-versa. At some level, it is misleading to consider spectrally-classed populations as representing two independent components of the galaxy distribution. Nonetheless, having established some form of evolution, it is important to find which kinds of sources are involved. A specific advantage of considering the strong [OII] galaxies is the high redshift completeness assured by their emission spectrum. This point was considered quantitatively by Colless et al. (1990) for continuum S/N-limited samples. In view of previous claims for the central role of star formation in understanding the counts, an '[OII] strong' subsample is likely to be a valuable dataset for analysis.

Figure 14 shows the change with redshift in the absolute magnitude distribution and derived LFs for galaxies separated according to whether their $W_\lambda$[OII] exceeds or is less than 20Å. Significant evolution of the $W_\lambda$[OII] the star-forming galaxies. These sources show evolution qualitatively similar, though noticeably stronger, than that observed for the whole population (cf. Table 3). Direct comparison of the high- and low-redshift LFs in Figure 14a shows the space density of star-forming galaxies has decreased at all luminosities by almost a factor of 2 between $z \simeq 0.4$ and $z \simeq 0.15$. This decline corresponds to an overall fading of the star-forming population of 0.5 mag over this redshift range.

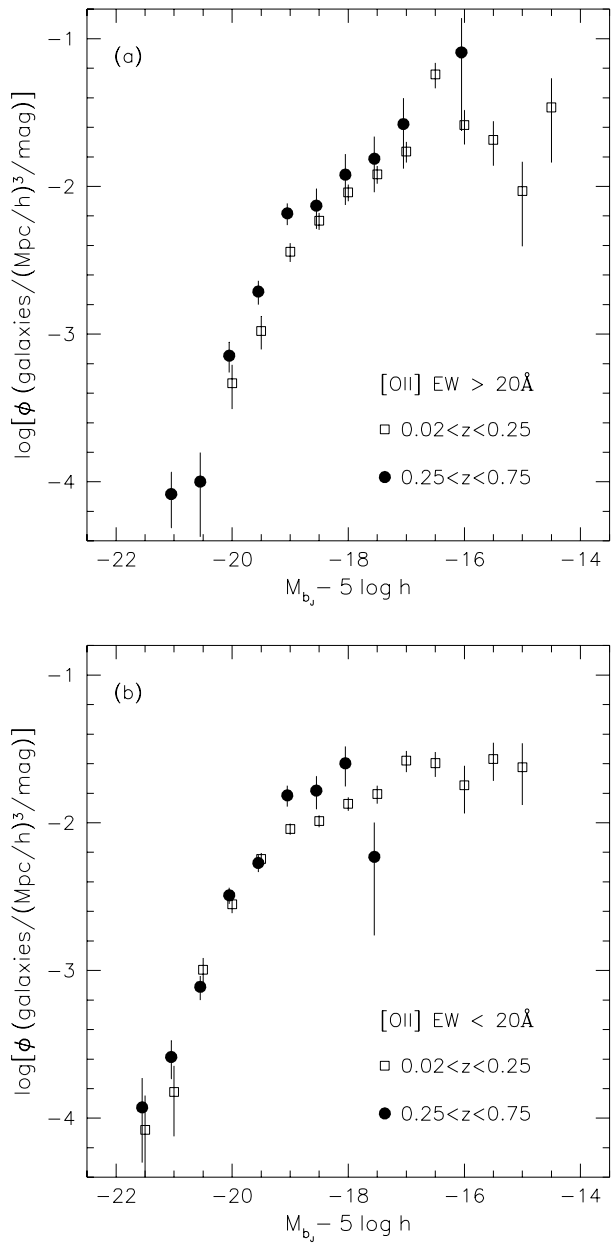

**Figure 14.** Luminosity functions at various redshifts for galaxies selected according to the equivalent width $W_\lambda$ of the [OII] 3727 Å emission line.

The rapid evolution in the LF of the star forming galaxies is consistent with a conclusion derived by Lilly et al. (1995) from the $I$-band selected CFRS survey. Those workers claim substantial brightening with redshift for $I$-selected galaxies whose rest-frame colours are bluer than Sbc type. In Paper II we address the question of evolution as a function of spectral class more rigorously. However, we note that Lilly et al.'s Figure 3(b) is quite similar to our Figure 14 which is particularly encouraging considering the different selection criteria and methods used by the two groups.

However, it should be noted that (i) by virtue of our $B$-selection we have a much greater sensitivity to this evolutionary trend, and (ii) the wider range in apparent magnitude surveyed here provides a clearer estimate of the trends with luminosity at each redshift. In the CFRS survey, which spans only a limited apparent magnitude range with its single magnitude-limited sample, it is very difficult to estimate the *shape* of the LF at any particular redshift. Conceivably this is why Lilly et al. are unable, from their LF results, to reproduce the $B$-band counts (their Fig. 8). On the other hand, the CFRS survey provides valuable information at higher redshifts (150 blue galaxies have $0.75 < z < 1.3$) by virtue of the reduced $k$-correction in the longer wavelength band. The two surveys therefore complement each other remarkably well.

The question arises as to whether the [OII] sources are simply fading as a separate self-contained population? The answer is unclear because it is, necessarily, somewhat of an arbitrary distinction as to whether a galaxy is put in the [OII] -strong or quiescent sample. Certainly, if one starts with the high redshift LF for the [OII] -strong galaxies in Fig. 14 and applies a progressive luminosity-independent fading with redshift to the entire population, it is not possible to reconstruct Fig. 8 at $z=0$. Without some form of differential fading, there are too many star-forming galaxies at high $z$ for the local representatives to be dimmed versions. No doubt the same dilemma would arise if one characterised the populations on the basis of colours as in Lilly et al. (1995). Neither colour nor spectral types is a particularly good classifier over a range in redshift since, when star formation falls below some threshold, a galaxy can easily change from one category to another. One possible way forward may be to use HST morphology as the basic classifier,although it will be some time before such sizeable samples are available (Glazebrook et al. 1995c, Ellis 1995).

## 5 CONCLUSIONS

We summarise our principal conclusions as follows:

(i) We have completed a major new redshift survey of 1026 galaxies at intermediate magnitude which, together with earlier published data secured by our team, allows us to construct a catalogue of over 1700 galaxy redshifts spanning a wide range in apparent magnitude from $b_J$=11.5 to 24. The wide range in implied luminosity is a significant step forward in determining directly the form of the luminosity function (LF) at various redshifts.

(ii) We confirm that the local LF has a Schechter faint end slope with $\alpha \simeq$-1.1 as claimed by Efstathiou et al. (1988) and Loveday et al. (1992). A significantly steeper slope would lead to the detection of many more low redshift galaxies than observed in the faintest surveys. A careful analysis of the local LF derived from catalogues limited at different apparent magnitudes shows the principal uncertainty in the local LF lies in its absolute normalisation not its shape. We present convincing evidence for a higher LF normalisation than that previously estimated, and this normalisation is in agreement with other, indirect, estimates recently published.

**Table 3.** Luminosity function fits as a function of [O II] equivalent width

| $W_\lambda$[OII] and $z$ range | $M^*$ | $\alpha$ | $\log \phi^*$ | $\phi^*$ | $\chi^2\ (\nu)$ |
|---|---|---|---|---|---|
| $W_\lambda >20$Å, $z<0.25$ | -18.42 [-0.14,+0.14] | -1.04 [-0.08,+0.10] | -1.70 [-0.06,+0.08] | 0.0200 | 22.53 ( 9) |
| $W_\lambda >20$Å, $z>0.25$ | -18.96 [-0.30,+0.32] | -1.44 [-0.26,+0.38] | -1.88 [-0.24,+0.20] | 0.0132 | 8.82 ( 7) |
| $W_\lambda <20$Å, $z<0.25$ | -19.42 [-0.12,+0.08] | -1.12 [-0.06,+0.04] | -1.76 [-0.06,+0.04] | 0.0174 | 9.59 (11) |
| $W_\lambda <20$Å, $z>0.25$ | -19.08 [-0.18,+0.16] | -0.74 [-0.22,+0.24] | -1.58 [-0.08,+0.08] | 0.0263 | 20.84 ( 6) |

(iii) Analysis of the galaxy LF as a function of redshift shows evidence for a steepening of the faint end slope with increasing redshift, from Schechter values of $\alpha=$-1.1 locally to $\alpha=$-1.5 at redshift $z \simeq 0.5$. There is also a marked increase in the number of $L^*$ galaxies over the look-back times sampled. We demonstrate the robustness of these results to various incompleteness effects inherent in the survey. These trends we have found provide a consistent explanation for the original puzzle of the excess galaxy counts and lack of evolution in the redshift distribution. The explanation confirms the original suggestion made by Broadhurst et al. (1988).

(iv) The evolution is consistent, to a reasonable approximation, with that arising primarily in the LF of strong star-forming galaxies categorised via the rest-frame equivalent width of [O II]3727 Å. There has been a decline by almost a factor of 2 in the mean luminosity density of these star-forming sources since $z \simeq 0.5$, and it is this evolution which is responsible for the faint blue galaxy excess.

(v) In common with recent conclusions derived from counts categorised by HST morphologies, our LF studies have highlighted two galaxy populations which evolve in very different ways. Massive galaxies at the bright end of the LF show only marginal changes in their rest-frame $B$ luminosities at recent times, whereas lower mass galaxies suffer a rapidly-declining star formation rate.

## ACKNOWLEDGEMENTS

We thank Ian Parry, Ray Sharples, Peter Gray and staff at the AAT for their hard efforts in making the Autofib instrument a reliable working system. We also thank PATT for their patience and support over many semesters. Keith Taylor and Jeremy Allington-Smith are thanked for their assistance with LDSS-1 and LDSS-2. We acknowledge useful discussions with Len Cowie, Olivier LeFevre, David Koo, Simon Lilly, Steve Maddox and Ron Marzke. RSE and KGB acknowledge financial support from PPARC. JSH thanks the Marshall Aid Commission. MMC acknowledges the assistance of the Australian Academy of Science/Royal Society exchange program.

16  *R.S.Ellis et al.*

King C.R. & Ellis R.S., 1985, ApJ, 288, 456
Kirschner R.P., Oemler A. & Schechter P., 1978, AJ, 83, 1549
Koo D.C. & Kron R., 1992, ARA&A, 30, 613
Koo D.C., Gronwall, C. & Bruzual, G.A. 1993, ApJ, 415, L21
Kron R., 1980, Vistas in Astronomy, 26, 37
Lacey C., Guiderdoni B., Rocca-Volmerange B. & Silk J., 1992, ApJ, 402, 15
Lilly S.J., Cowie L.L. & Gardner J.P. 1991, ApJ, 369, 79
Lilly S.J., 1993, ApJ, 411, 501
Lilly S.J., Tresse L., Hammer F., Crampton, D. & LeFevre, O. 1995, ApJ, in press.
Loveday J., Peterson B.A., Efstathiou G. & Maddox S.J., 1992, ApJ, 390, 338
McGaugh S., 1994, Nature, 367, 538
Maddox S.J., Sutherland W.J., Efstathiou G.P., Loveday J. & Peterson B.A., 1990, MNRAS 273, 257
Marzke R., Huchra, J.P. & Geller, M, 1994, ApJ 428, 43
Metcalfe N., Shanks T., Fong R. & Roche, N. 1995a, MNRAS, 273, 257.
Metcalfe N., Fong R. & Shanks,T. 1995b, MNRAS, 274, 769.
Mutz S.B. et al., 1994, ApJL, 434, L55
Parry I.R. & Sharples R.M. 1988, in *Fiber Optics*, ed. Barden, S.M., PASP Conference Series, 3, 93
Pence W., 1976, ApJ, 188, 444
Peterson B.A., Ellis R.S., Bean A.J., Efstathiou G.P., Shanks T., Fong R. & Zou Z-L., 1985, MNRAS, 221, 233
Phillips A.C. et al., 1995, ApJ, in press
Phillipps S. & Driver S., 1995, MNRAS, in press
Roukema, B. & Peterson, B.A. 1995, ApJ, in press
Saunders W., Rowan-Robinson M., Lawrence A., Efstathiou G., Kaiser N., Ellis R.S. & Frenk C.S., 1990, MNRAS, 242, 318
Schechter P., 1976, ApJ, 203, 297
Schmidt M., 1968, ApJ, 151, 393
Schwartzenberg J.M., Phillipps S., Smith R.M., Couch W.J. & Boyle B.J., MNRAS, in press
Steidel C., Dickinson, M. & Persson E., 1995, ApJ, in press.
Tritton K.P. & Morton D.C., 1984, MNRAS, 209, 429